\documentclass[aps,prd,showpacs,preprintnumbers,amsmath,amssymb,nofootinbib]{revtex4}
\usepackage{graphicx}
\usepackage{verbatim}
\usepackage{slashed}
\usepackage{amsthm}
\usepackage{amsfonts}
\usepackage[dvips]{color}

\definecolor{gray}{RGB}{200,200,200}

\def\ththMat#1#2#3#4#5#6#7#8#9{\ensuremath{\begin{pmatrix}#1&#2&#3\\
                                                          #4&#5&#6\\
                                                          #7&#8&#9\end{pmatrix}}}
\def\twoneMat#1#2{\ensuremath{\begin{pmatrix}#1\\
                                              #2\end{pmatrix}}}
\def\twtwMat#1#2#3#4{\ensuremath{\begin{pmatrix}#1&#2\\
                                                #3&#4\end{pmatrix}}}
\def\onetwMat#1#2{\ensuremath{\begin{pmatrix}#1&#2\end{pmatrix}}}

\def\order#1{\ensuremath{\mathcal{O}\left(#1\right)}}

\def\BR#1{\ensuremath{\text{Br}(#1)}}
\def\drate#1{\ensuremath{\Gamma(#1)}}

\begin{document}

\title{Constraints on exotic lepton doublets with minimal coupling to the standard model}
\author{Sandy~S.~C.~Law}
\email{slaw@mail.ncku.edu.tw}
\affiliation{Department of Physics, Chung Yuan Christian University,
Chung-Li, Taiwan 320, Republic of China.}

\date{5 February, 2011}
\preprint{CYCU-HEP-11-07}
\pacs{12.60.-i, 13.40.Em, 14.60.Hi}

\begin{abstract}
We investigate the consequences of introducing a set of exotic doublet leptons which couple to the standard model leptons in a minimal way. Through these additional gauge invariant and renormalizable coupling terms, new sources of tree-level flavor changing currents are induced via mixing. In this work, we derive constraints on the parameters that govern the couplings to the exotic doublets by invoking the current low-energy experimental data on processes such as leptonic $Z$ decays, $\ell \rightarrow 3 \ell'$,  $\ell \rightarrow \ell' \gamma$, and $\mu$-$e$ conversion in atomic nuclei. Moreover, we have analyzed the role  these doublets play on the lepton anomalous magnetic moments, and found that their contribution is negligible.
\end{abstract}

\maketitle


\section{Introduction}

It is well-known that the phenomenon of neutrino oscillations \cite{neutrinos_exp} have motivated an extensive study on models with nonzero neutrino masses and lepton flavor violation (LFV). Central to all these investigations is the introduction of new interactions (and most likely new particles too) to the minimal standard model (SM). While there can be many different ways of inventing and restricting the new physics, recently in \cite{Chua:2010me}, a fairly economical approach based on only SM gauge invariance, renormalizability and the concept of ``minimal coupling'' was considered for the lepton sector.

In that work, a generic minimal coupling between SM particles and some exotic field was defined to have the form
\begin{equation}\label{eqn:general_form}
 Y_\text{exotic} \;\text{(SM particle)}\cdot \text{(SM particle)} \cdot \text{(exotic particle)}\;,
\end{equation}
where $Y_\text{exotic}$ denotes the coupling strength and the ``exotic particle'' can either be a scalar boson, a fermion or a vector boson. All particles in (\ref{eqn:general_form}) are assumed to be \emph{uncolored} in the $SU(3)_c$ sense because we only wish to extend the lepton sector. These minimal interactions are of interest because they are relatively simple and their collider signatures  may be detected at the LHC in the near future under favorable conditions \cite{DelNobile:2009st}.

Schematically, there are five distinct types of interaction with the SM fields allowed by (\ref{eqn:general_form}) and the requirement of renormalizability\footnote{We do not consider terms such as (SM Higgs)(SM Higgs)(new boson) where no leptons of any type is present.},
\begin{eqnarray}\label{equ:5_types}
&\text{(i)}\;\;
 L_L \times L_L \times [\text{new}]\;,
 \;\;
\text{(ii)}\;\;
 L_L \times \ell_R \times [\text{new}]\;,
  \;\;
\text{(iii)}\;\;
 \ell_R \times \ell_R \times [\text{new}]\;, \nonumber \\
&\text{(iv)}\;\;
 L_L \times \phi \times [\text{new}]\;,
 \;\;
\text{(v)}\;\;
 \ell_R \times \phi \times [\text{new}] \;.
\end{eqnarray}
where $L_L =  (\nu_L, \ell_L)^T$ is the left-handed (LH) lepton doublet, $\ell_R$ is the right-handed (RH) lepton singlet, and $\phi=(\phi^+, \phi^0)^T$ denotes the SM Higgs doublet. The imposition of Lorentz and SM gauge symmetries will then result in only 13 types of exotic multiplets that are minimally coupled to the SM particles (see Table~\ref{table:field_summary}).

As pointed out in  \cite{Chua:2010me}, it seems that all but \emph{two} types of the new particles induced by this setup has already been heavily analyzed because of various motivations. The two that are rarely discussed are the exotic lepton triplets $E_{R,L}=(E^0_{L,R},E^{-}_{L,R},E^{--}_{L,R})^T$ and the doublets $\widetilde{L}_{L,R} = (\widetilde{L}^-_{L,R},\widetilde{L}^{--}_{L,R})^T$ with SM quantum numbers $(1,3,-1)$ and $(1,2,-3/2)$ respectively.
Whereas the study of the triplets $E_{R,L}$ formed the backbone of \cite{Chua:2010me}, we shall concentrate on the last remaining possibility---doublets $\widetilde{L}_{L,R}$ in this work.

The various implications of introducing $\widetilde{L}_{L,R}$ to the SM are studied in the subsequent sections. Processes such as $Z$ decays (Sec.~\ref{sec:Z_decays}), LFV decays: $\ell\rightarrow 3\ell'$ (Sec.~\ref{sec:L23L}), $\ell\rightarrow \ell' \gamma$ (Sec.~\ref{sec:L2eg}), and $\mu$-$e$ conversion in atomic nuclei (Sec.~\ref{sec:m_to_e}) are considered with the aim to derive constraints on the relevant new physics parameters using low-energy experimental data (Sec.~\ref{sec:global_fit}).
The contributions from $\widetilde{L}_{L,R}$ to the lepton anomalous magnetic moments will be investigated in Sec.~\ref{sec:AMM}. Finally, in Sec.~\ref{sec:LHC}, we will make some brief comments regarding their collider phenomenologies.

\begin{table}[ht]
\begin{center}
\begin{tabular}{|c|c|c|c|c|c|c|}
\hline
[new] & spin & $SU(2)_L$ & $U(1)_Y$ &  type & SM fields involved & studied in\\
\hline
$\Phi_i$ & 0 & 2 & $1/2$ & (ii) & $\overline{L}_L\;\; \ell_R$ & multi-Higgs doublet models \cite{multi_Higgs_models, rad_seesaw_eg}\\
$\chi_0$ & 0 & 1 & $-1$ &  (i) &$\overline{L}_L\;\; L_L^c$ & dilepton/Babu-Zee models \cite{rad_seesaw_eg, dileptons1, dileptons2, babu_zee}\\
$\Delta$ & 0 & 3 & $-1$ &  (i) &$\overline{L}_L\;\; L_L^c$ & dilepton/Type-II seesaw
                                   \cite{dileptons1, doubly_Higg, type2seesaw, Abada:2007ux}\\
$\xi_0$ & 0 & 1 & 2 &  (iii) &$\overline{\ell}_R\;\; \ell_R^c$ & dilepton/Babu-Zee models \cite{dileptons1, dileptons2, babu_zee, babu_zee2}
\\%
\hline
$\nu_R$ & 1/2 & 1 & $0$ &  (iv) &$\overline{L}_L\;\; \phi^c$ & Type-I seesaw \cite{Abada:2007ux, type1seesaw, Biggio:2008in, type1_3, type1set2}\\
$\Sigma_R$ & 1/2 & 3 & $0$ &  (iv) &$\overline{L}_L\;\; \phi^c$ & Type-III seesaw \cite{Abada:2007ux, type1_3, type3seesaw, Abada:2008ea}\\
$L_L''$ & 1/2 & 2 & $-1/2$ &  (v) &$\overline{\ell}_R\;\; \phi^\dagger$ & 4th generation leptons \cite{4thgen}\\
$\ell_R''$ & 1/2 & 1 & $-1$ & (iv) &$\overline{L}_L\;\; \phi$ & 4th generation leptons \cite{4thgen}\\
$E_{R,L}$ & 1/2 & 3 & $-1$ & (iv) &$\overline{L}_L\, \phi\;$  ($E_R$ only) & see \cite{Chua:2010me, Delgado:2011iz}\\
$\widetilde{L}_{L,R}$ & 1/2 & 2 & $-3/2$ &  (v) & $\overline{\ell}_R \,(\phi^{c})^{\dagger}$ ($\widetilde{L}_L$ only)&  \emph{rarely discussed} \\
\hline
$Z_{\mu}'$ & 1 & 1 & $0$ & (i) \& (iii) & $\overline{L}_L\;\; L_L\;$ and $\;\overline{\ell}_R\;\; \ell_R$& \\
$X_{\mu}$ & 1 & 2 & $-3/2$ & (ii) & $\overline{L}_L\;\; \ell_R^c$& GUT/dilepton boson models \cite{dileptons1, dilepton_boson}\\
$W_{\mu}'$ & 1 & 3 & $0$ & (i) &$\overline{L}_L\;\; L_L$ & \\
\hline
\end{tabular}
\caption{
Summary of the 13 types of exotic multiplets induced by the 5 general types of minimal couplings:
(i) $L_L \times L_L \times [\text{new}]$,
(ii) $L_L \times \ell_R \times [\text{new}]$,
(iii) $\ell_R \times \ell_R \times [\text{new}]$,
(iv) $L_L \times \phi \times [\text{new}]$,
and
(v) $\ell_R \times \phi \times [\text{new}]$.
Hypercharges are defined with $Q=I_3+Y$.}
\label{table:field_summary}
\end{center}
\end{table}

\section{Model with exotic lepton doublets, $\widetilde{L}_{L,R}$}\label{sec:the_model}

We begin by writing down the framework and explaining our notations for our minimally extended SM with exotic lepton doublets. In this model, we have two new sets (LH and RH) of lepton doublets $\widetilde{L}_{L,R}$, all having hypercharge $-3/2$ (where $Q=I_3+Y$). In matrix form, they are given by
\begin{equation}
 \widetilde{L}_L =\twoneMat{\widetilde{L}^{-}_L}{\widetilde{L}^{--}_L}\;,
 \qquad
 \widetilde{L}_R =\twoneMat{\widetilde{L}^{-}_R}{\widetilde{L}^{--}_R}\;
 \sim (1,2,-3/2)\;,
\end{equation}
where $\widetilde{L}_L$ and $\widetilde{L}_R$ are independent fields.\footnote{Because of the identical transformation properties for LH and RH fields, chiral anomalies cancel automatically in this setup. Furthermore, by introducing a pair of these, we have maintained an even number of doublets in the overall model, and hence, avoiding any issues with global $SU(2)$ anomalies \cite{Witten:1982fp}.}
The interaction Lagrangian of interest is\footnote{We have also introduced three RH neutrino fields, $\nu_R$, so that neutrinos can have a Dirac mass. This is done so because we know that neutrinos are massive. However, to avoid making the model too complicated and the risk of masking the effects from the exotic doublets, we have opted \emph{not} to include a Majorana mass term for simplicity. But we shall briefly comment on the effects of the Majorana mass term at the end of this section. For a full discussion though the readers may refer to the work of \cite{Abada:2007ux, Antusch:2006vwa}.}
\begin{equation}\label{equ:main_L}
 \widetilde{\mathcal{L}} =
 \overline{\widetilde{L}_L} i\slashed{D} \widetilde{L}_L +
 \overline{\widetilde{L}_R} i\slashed{D} \widetilde{L}_R
  -\left[\overline{\widetilde{L}_L} \widetilde{M} \widetilde{L}_R
  +  \overline{\widetilde{L}_L}\widetilde{Y}^\dagger \phi^c \ell_R
  +  \overline{L}_L Y_\ell \phi \,\ell_R
  +  \overline{L}_L Y_\nu \phi^c \,\nu_R
  + h.c.
  \right]\;,
\end{equation}
where $\phi^c = (\phi^{0*}, -\phi^-)^T$, and the covariant derivative is defined as,
\begin{equation}
\slashed{D} = \slashed{\partial}
               -\frac{ig}{\sqrt{2}} \left[\slashed{W}^+ \twtwMat{0}{1}{0}{0}
               +
               \slashed{W}^- \twtwMat{0}{0}{1}{0}
               \right]
               -\frac{ig}{\cos \theta_w} \slashed{Z} (I_3 - \sin^2 \theta_w Q)
               +i e \slashed{A} Q\;, \quad (e>0)\;,
\end{equation}
with $Q$ and $I_3$ being the operators for electric charge and the 3rd component of isospin respectively.
In (\ref{equ:main_L}), the Yukawa term involving $\widetilde{Y}^\dagger$ defines the minimal coupling between $\ell_R$ and $\widetilde{L}_L$ while $\widetilde{M}$ sets the energy scale of the new physics.
Notice that one cannot have a similar type of minimal coupling between $\widetilde{L}_R$ and any other SM leptons because of SM gauge invariance. But its effects on the SM sector can enter indirectly via the mass term $\overline{\widetilde{L}_L} \widetilde{M} \widetilde{L}_R$.
Writing out all the relevant interactions in (\ref{equ:main_L}), we have
\begin{equation}
 \widetilde{\mathcal{L}} +\mathcal{L}^\text{SM} =
 \mathcal{L}^W+
 \mathcal{L}^Z+
 \mathcal{L}^\text{mass}+
 \mathcal{L}^H
 + \cdots
 \;,
\end{equation}
where
\begin{align}
 \mathcal{L}^W &=
  \frac{g}{\sqrt{2}} \left[
   \overline{\nu}_L \slashed{W}^+ \ell_L
    +
     \overline{\widetilde{L}^{-}_L} \slashed{W}^+ \widetilde{L}^{--}_L
    +\overline{\widetilde{L}^{-}_R} \slashed{W}^+ \widetilde{L}^{--}_R
  \right]
   + h.c.\;, \label{equ:L_W}\\
  \mathcal{L}^Z &=
  \frac{g}{\cos \theta_w} \left[
   \frac{1}{2}\; \overline{\nu}_L \slashed{Z} \nu_L
   +\left(-\frac{1}{2} +  \sin^2 \theta_w\right) \overline{\ell}_L \slashed{Z} \ell_L
   +\sin^2 \theta_w \;\overline{\ell}_R \slashed{Z} \ell_R
   +\left(\frac{1}{2}+\sin^2\theta_w\right) \overline{\widetilde{L}^{-}_L} \slashed{Z} \widetilde{L}^{-}_L
   \right.\nonumber\\
   &\qquad\qquad\quad\left.
   +\left(\frac{1}{2}+\sin^2\theta_w\right) \overline{\widetilde{L}^{-}_R} \slashed{Z} \widetilde{L}^{-}_R
   +\left(-\frac{1}{2}+2\sin^2\theta_w\right) \overline{\widetilde{L}^{--}_L} \slashed{Z} \widetilde{L}^{--}_L
   +\left(-\frac{1}{2}+2\sin^2\theta_w\right) \overline{\widetilde{L}^{--}_R} \slashed{Z} \widetilde{L}^{--}_R
   \right]\;,\label{equ:L_Z}\\
  \mathcal{L}^\text{mass} &=
  -\overline{\widetilde{L}^{-}_L} \widetilde{M} \widetilde{L}^{-}_R
  -\overline{\widetilde{L}^{--}_L} \widetilde{M} \widetilde{L}^{--}_R
  -\frac{v}{\sqrt{2}} \overline{\widetilde{L}^{-}_L} \,\widetilde{Y}^\dagger\, \ell_R
  -\overline{\ell}_L m_\ell \ell_R
  -\overline{\nu}_L m_D \nu_R + h.c.\;, \label{equ:L_mass}\\
  \mathcal{L}^H &=
  -\frac{1}{\sqrt{2}} \overline{\widetilde{L}^{-}_L}\widetilde{Y}^\dagger \ell_R H
  -\frac{1}{v}\overline{\ell}_L m_\ell \ell_R H
  -\frac{1}{v}\overline{\nu}_L m_D \nu_R H + h.c.\;. \label{equ:L_H}
\end{align}
In getting (\ref{equ:L_mass}) and (\ref{equ:L_H}), we have written $\phi=(\phi^+, \phi^0)^T \equiv (\phi^+, (v + H + i\eta)/\sqrt{2})^T$, where $v$ is the Higgs vacuum expectation value (chosen to be real), $\eta$ and $\phi^{\pm}$ are the would-be Goldstone bosons. Also, we have defined $m_\ell \equiv v Y_\ell/\sqrt{2}$ and $m_D\equiv v Y_\nu/\sqrt{2}$.

The mixing between the SM charged leptons and components of the exotic doublet may be readily derived if we write the relevant terms in $\mathcal{L}^\text{mass}$ in matrix form:
\begin{align}
\mathcal{L}^\text{mass} &=
 -\frac{1}{2}\onetwMat{\overline{\nu}_L}{\overline{(\nu_R)^c}}
  \twtwMat{m_D}{0}{0}{m_D^T}
   \twoneMat{\nu_R}{(\nu_L)^c}
 -\onetwMat{\overline{\ell}_L}{\overline{\widetilde{L}^{-}_L}}
  \twtwMat{m_\ell}{0}{v \widetilde{Y}^\dagger/\sqrt{2}}{\widetilde{M}}
   \twoneMat{\ell_R}{\widetilde{L}^{-}_R}
\nonumber \\
 &\qquad
 -\frac{1}{2}\onetwMat{\overline{\widetilde{L}^{--}_L}}{\overline{(\widetilde{L}^{--}_R)^c}}
  \twtwMat{\widetilde{M}}{0}{0}{\widetilde{M}^T}
   \twoneMat{\widetilde{L}^{--}_R}{(\widetilde{L}^{--}_L)^c}
   + \cdots\;, \label{equ:L_mass_matrix}
\end{align}
where we have used the fact that $\overline{\psi_1}\psi_2 \equiv \overline{\psi_2^c}\psi_1^c$ for any fermion field $\psi_{1,2}$. One may define the following unitary transformations to bring all fields into their mass eigenbasis:
\begin{equation} \label{equ:UV_trans}
 \twoneMat{\nu_{L,R}}{(\nu_{R,L})^c}
 = V_{L,R}
 \twoneMat{\nu_{L,R}}{(\nu_{R,L})^c}_m\;,
 \quad
 \twoneMat{\ell_{L,R}}{\widetilde{L}_{L,R}^-}
 = U_{L,R}
 \twoneMat{\ell_{L,R}}{\widetilde{L}_{L,R}^-}_m \;,
  \quad
 \twoneMat{\widetilde{L}^{--}_{L,R}}{(\widetilde{L}^{--}_{R,L})^c}
 = I_n
 \twoneMat{\widetilde{L}^{--}_{L,R}}{(\widetilde{L}^{--}_{R,L})^c}_m \;,
\end{equation}
where the subscript $m$ indicates the mass basis. In (\ref{equ:UV_trans}), $V_{L,R}, U_{L,R}$ and $I_n$ have dimensions $6\times6, (3+n)\times (3+n)$ and $n\times n$ respectively, with $n$ being the number of generations of the exotic $\widetilde{L}_{L,R}$ fields added to the model.

Without loss of generality, one can choose to work in the basis where $m_\ell$ and $\widetilde{M}$ are real and diagonal. As a result, $I_n$ is in fact the identity matrix. Furthermore, to make the notations less cluttered, we absorb  the neutrino right diagonalization matrix $U_{\nu R}$ into $m_D$. In other words, we set $m_D \equiv m_D' U_{\nu R}$, where $U_{\nu L}^\dagger m_D' U_{\nu R} = m_D^\text{diag}$ and $U_{\nu L}$ is the neutrino left diagonalization matrix. With these conventions, and to $\order{v^2 \widetilde{M}^{-2}}$, the transformation matrices are given by
\begin{align}
 V_L = \twtwMat{U_{\nu L}}{0}
             {0}{1} \;,
 &\qquad
 V_R = \twtwMat{1}{0}{0}{U_{\nu L}^*}
  \;,\label{equ:V_trans}\\
  U_L = \twtwMat{1}{v\, m_\ell \widetilde{Y} \widetilde{M}^{-2}/\sqrt{2}}{-v \widetilde{M}^{-2} \widetilde{Y}^\dagger m_\ell /\sqrt{2}}{1}
 \;,
 &\qquad
 U_R = \twtwMat{1-\lambda}{v \widetilde{Y} \widetilde{M}^{-1}/\sqrt{2}}{-v \widetilde{M}^{-1} \widetilde{Y}^\dagger /\sqrt{2}}{1-\lambda'} \;,
\label{equ:U_trans}
\end{align}
where
\begin{equation}\label{eqn:def_lambda}
 \lambda \equiv \frac{v^2}{2} \,\widetilde{Y}\, \widetilde{M}^{-2}\, \widetilde{Y}^\dagger \;,
 \quad \text{ and }\quad
 \lambda' \equiv \frac{v^2}{2} \widetilde{M}^{-1}\, \widetilde{Y}^\dagger\, \widetilde{Y} \widetilde{M}^{-1}\,\;,
\end{equation}
are $3\times 3$ and $n\times n$ matrices in flavor space respectively. Note that $U_{\nu L}$ may be identified as the usual neutrino mixing matrix, $U_\text{PMNS}$.

The relevant terms in the interaction Lagrangian with respect to the mass eigenbasis are therefore given by
\begin{align}
 \mathcal{L}^W &= \frac{g}{\sqrt{2}} \left[
  \onetwMat{\overline{\nu}}{\overline{\nu^c}}_m \slashed{W}^+
   \left[P_L \,g_{1L}^{W} +P_R \,g_{1R}^{W}\right]
   \twoneMat{\ell}{\widetilde{L}^-}_m
   +
   \onetwMat{\overline{\ell}}{\overline{\widetilde{L}^{-}}}_m \slashed{W}^+
   \left[P_L \,g_{2L}^{W} +P_R \,g_{2R}^{W}\right]
   \twoneMat{\widetilde{L}^{--}}{(\widetilde{L}^{--})^c}_m
   \right] +h.c.
    , \label{equ:L_CC} \\
  \mathcal{L}^Z &= \frac{g}{\cos \theta_w}
  \left[
     \onetwMat{\overline{\nu}}{\overline{\nu^c}}_m \slashed{Z}
   \,P_L \,g_{1L}^{Z}
   \twoneMat{\nu}{\nu^c}_m
   +
    \onetwMat{\overline{\ell}}{\overline{\widetilde{L}^{-}}}_m \slashed{Z}
   \left[P_L \,g_{2L}^{Z} +P_R \,g_{2R}^{Z}\right]
   \twoneMat{\ell}{\widetilde{L}^{-}}_m
   \right.\nonumber
   \\
   &\qquad\qquad\qquad\left.
 + \onetwMat{\overline{\widetilde{L}^{--}}}{\overline{(\widetilde{L}^{--})^c}}_m \slashed{Z}
   \,P_L \,g_{3L}^{Z}
   \twoneMat{\widetilde{L}^{--}}{(\widetilde{L}^{--})^c}_m
   \right]
   \;,  \label{equ:L_NC}\\
   \mathcal{L}^H &=
     \onetwMat{\overline{\nu}}{\overline{\nu^c}}_m H
   \,P_R \,g_{1R}^{H}\,
   \twoneMat{\nu}{\nu^c}_m
   +
     \onetwMat{\overline{\ell}}{\overline{\widetilde{L}^{-}}}_m H
   \left[P_L \,g_{2L}^{H} + P_R \,g_{2R}^{H}\right]
   \twoneMat{\ell}{\widetilde{L}^{-}}_m
   \label{equ:L_higgs}\;,
\end{align}
with the new generalized coupling matrices given by (to leading order)
\begin{align}
 g_{1L}^{W} &=
 \twtwMat{U_{\nu L}^\dagger}{v\, U_{\nu L}^\dagger m_\ell \widetilde{Y}\widetilde{M}^{-2}/\sqrt{2}}{0}{0} \;,
 \quad
 g_{2L}^{W} =
 \twtwMat{-v\, m_\ell \widetilde{Y}\widetilde{M}^{-2}/\sqrt{2}}{0}{1}{0} \;,
 \quad
  g_{1R}^{W} = 0 \;,
 \quad
 g_{2R}^{W} =  \twtwMat{-v\, \widetilde{Y}\widetilde{M}^{-1}/\sqrt{2}}{0}{1-\lambda'}{0}
 \;,
 \label{equ:g_W}\\
%
 g_{1L}^{Z} &= \twtwMat{1/2}{0}{0}{0} \;, \;
 g_{2L}^{Z} = \twtwMat{-1/2+\sin^2 \theta_w}{-v\, m_\ell \widetilde{Y}\widetilde{M}^{-2}/\sqrt{2}}
 {-v\, \widetilde{M}^{-2} \widetilde{Y}^\dagger m_\ell/\sqrt{2}}{1/2+\sin^2 \theta_w} \,, \;
 g_{3L}^{Z} = \twtwMat{-1/2+2\sin^2 \theta_w}{0}{0}{1/2-2\sin^2 \theta_w} \;,    \nonumber\\
 g_{2R}^{Z} &= \twtwMat{\sin^2 \theta_w +(1/2-\sin^2 \theta_w)\lambda}
 {-v\, \widetilde{Y}\widetilde{M}^{-1}/2\sqrt{2}}
 {-v\,\widetilde{M}^{-1} \widetilde{Y}^\dagger/2\sqrt{2}}
 {1/2+\sin^2 \theta_w -(1+\sin^2 \theta_w)\lambda'} \;, \label{equ:g_Z}\\
%
 g_{1R}^{H} &= \twtwMat{-U_{\nu L}^\dagger m_D/v}{0}{0}{0} \;,\;
 g_{2R}^{H} = \twtwMat{m_\ell(2\lambda-1)/v}{-m_\ell\,\widetilde{Y}\widetilde{M}^{-1}/\sqrt{2}}
 {\widetilde{Y}^\dagger (\lambda -1)/\sqrt{2} -\widetilde{M}^{-2}\widetilde{Y}^\dagger m_\ell^2/\sqrt{2}\;}
 {-v\,\widetilde{Y}^\dagger\widetilde{Y}\widetilde{M}^{-1}/2} \;,
  \;
  g_{2L}^{H} =  \left(g_{2R}^{H}\right)^\dagger\;.
  \label{equ:g_H_both}
\end{align}
Note that there is no need to include $g_{1R}^{Z},g_{3R}^{Z}$ and $g_{1L}^{H}$ because they are redundant when the lepton fields are grouped in this matrix form.\footnote{We point out in passing that in \cite{Chua:2010me}, the corresponding coupling matrices which mix the ordinary charged leptons ($\ell$) with the doubly charged exotic particles (cf. $g_{2L,R}^{W}$ in this model) were \emph{not} investigated. While their omission would not affect the overall conclusions reached in \cite{Chua:2010me}, they do result in a slight change of the numerical values coming from analyzing the $\ell \rightarrow \ell' \gamma$ graphs. It is therefore more complete to include them as has been done here for this doublet model (see Sec.~\ref{sec:L2eg}).}
The setup described above will allow us to easily study the new phenomenologies and any subsequent constraints of introducing these exotic doublets.

Any new contributions to tree-level flavor changing currents will be provided by the nonzero off-diagonal entries of matrix $\lambda$. For instance, the presence of  $\lambda$ in the upper-left $(3\times 3)$-block of $g^{Z}_{2R}$ is indicative of this fact. But despite the introduction of new mixing effects in certain sectors of the theory, one observes that the SM charged current interaction remains unaltered at tree-level. This is understandable since the coupling to the $W$ boson is only non-trivial for LH  particles in the SM while the exotic fields connect to the RH sector exclusively. As a result, this also explains the reason $\lambda$ enters  only in the $P_R$ term of the SM neutral current Lagrangian (in index form):
\begin{align}
  \mathcal{L}^{NC} &= \frac{g}{\cos \theta_w}\left\{
  \overline{\nu_i} \, \slashed{Z}\, P_L
  \,\frac{\delta_{ij}}{2} \, \nu_j
  +
  \overline{\ell}_\alpha \, \slashed{Z}\,
  \left[
  \left(\sin^2\theta_w  -\frac{P_L}{2}\right)\delta_{\alpha\beta}
  +P_R 
  \left(\frac{1}{2}-\sin^2\theta_w \right)\lambda_{\alpha\beta}
  \right]\ell_\beta
  \right\}\;.
   \label{equ:L_NC_R}
\end{align}
A related observation regarding the differences between this model and the one studied in \cite{Chua:2010me} is that the modification to the value of the Fermi constant as extracted from muon decay experiments ($\mu \rightarrow e + \text{missing energy}$) will now enter at \order{\lambda^2}:
\begin{equation}
 (G_F')^2 = G_F^2 (1+ \mathcal{C}|\lambda_{e\mu}|^2)\,,
\end{equation}
where $G_F \equiv \sqrt{2}\,g^2/8 M_W^2$ and $G_F'$ denotes the new Fermi constant in the presence of the new physics invoked by the exotic doublet $\widetilde{L}$, while  $\mathcal{C}$ is a numerical factor of \order{1}. Hence, to a very good approximation, we may simply take $G_F' \simeq G_F$ (ie. $\lambda$ independent) in all our calculations.

\subsection*{Comments on the neutrino Majorana mass term}

In this subsection, we make a quick digression to comment on the effects one would get if the neutrino Majorana mass term, 
$-M_R\, \overline{(\nu_R)^c}\, \nu_R/2$, is included in Lagrangian (\ref{equ:main_L}). 
As it is well-known that such Majorana mass term will induce mixing between the fields $\nu_L$ and $\nu_R$, leading to light and heavy mass eigenstates. The amount of such mixing is characterized by the new physics scale $M_R$ (or more precisely the ratio, $Y_\nu/M_R$). We shall demonstrate below that this effect can eventuate in the modification of the interaction Lagrangian that is akin to the role played by $\lambda$ and $\lambda'$ from earlier.

For the following discussion, we shall assume that there are three generations of $\nu_R$ and (without loss of generality) that the $3\times 3$ mass matrix $M_R$ is real and diagonal. Upon including the Majorana mass for $\nu_R$ in (\ref{equ:main_L}), it would be more convenient to rewrite the first term in (\ref{equ:L_mass_matrix}) as
\begin{equation}\label{eqn:L_mass_mat2}
\mathcal{L}^\text{mass}_2 =
 -\frac{1}{2}\onetwMat{\overline{\nu}_L}{\overline{(\nu_R)^c}}
  \twtwMat{0}{m_D}{m_D^T}{M_R}
   \twoneMat{(\nu_L)^c}{\nu_R} +\cdots\;.
\end{equation}
As usual, one can turn the fields into mass eigenstates via unitary transformations:
\begin{equation}
  \twoneMat{\nu_{L}}{(\nu_{R})^c}
 = {V_L}'
 \twoneMat{\nu_{L}}{(\nu_{R})^c}_m\;,
 \quad
  \twoneMat{(\nu_{L})^c}{\nu_{R}}
 = ({V_L}')^*
 \twoneMat{(\nu_{L})^c}{\nu_{R}}_m\;,
\end{equation}
with ${V_L}'$ defined by
\begin{equation}\label{eqn:V_L_prime}
 {V_L}' = \twtwMat{1-\epsilon}{-m_D M_R^{-1}}
                  {M_R^{-1} m_D^\dagger}{1-\epsilon'}
           \twtwMat{U_{\nu L}}{0}
                   {0}{1}
        = \twtwMat{(1-\epsilon)U_{\nu L}}{-m_D M_R^{-1}}
                  {M_R^{-1} m_D^\dagger U_{\nu L}}{1-\epsilon'}\;,
\end{equation}
and
\begin{equation}
 \epsilon \equiv \frac{v^2}{2} \,Y_\nu\, M_R^{-2}\, Y_\nu^\dagger 
 =m_D M_R^{-2} m_D^\dagger
 \quad ;\quad
 \epsilon' \equiv \frac{v^2}{2} M_R^{-1}\, Y_\nu^\dagger\, Y_\nu M_R^{-1}
 =M_R^{-1} m_D^\dagger m_D M_R^{-1}
 \;.
\end{equation}
This transformation will lead to the typical type-I seesaw result for light neutrinos: $U_{\nu L} m_\nu U_{\nu L}^T \simeq -m_D M_R^{-1} m_D^T$. It is important to note that $\epsilon, \epsilon' \simeq \order{M_R^{-2}}$ can be comparable to $\lambda, \lambda' \simeq \mathcal{O}\,(\widetilde{M}^{-2})$ if the two new physics scales are similar. Hence, this is the reason we have introduced them in definition (\ref{eqn:V_L_prime}). Physically, the quantity $\epsilon$ represents the flavor mixing correction to the LH neutrino kinetic energy terms. Such effect is originated from a dim-6 gauge invariant operator in the effective Lagrangian as discussed in \cite{Antusch:2006vwa}. After spontaneous symmetry breaking, the dim-4 kinetic terms receives small contribution from this dim-6 operator, which then leads to a non-unitary mixing  matrix for ordinary leptons. This result is evident from the modified SM interaction Lagrangian for the charged and neutral currents after introducing $\widetilde{L}_{L,R}$ and $\nu_R$ (with Majorana mass):
\begin{align}
  \mathcal{L}^{CC}_2 &=  \frac{g}{\sqrt{2}}\,
   \overline{\nu_i}\, \slashed{W}^+ P_L
   \left[
    (U_{\nu L}^\dagger)_{i\alpha}
    (\delta_{\alpha\beta} - \epsilon_{\alpha\beta})
   \right] \ell_\beta + h.c.\;, \label{eqn:LCC_2}
   \\
  \mathcal{L}^{NC}_2 &= \frac{g}{\cos \theta_w}\left\{
  \overline{\nu_i} \, \slashed{Z}\, P_L
  \left(
  \frac{\delta_{ij}}{2} - (U_{\nu L}^\dagger)_{i\alpha}
  \epsilon_{\alpha\beta} (U_{\nu L})_{\beta j}
  \right) \nu_j
  +
  \overline{\ell}_\alpha \, \slashed{Z}\,
  \left[
  \left(\sin^2\theta_w  -\frac{P_L}{2}\right)\delta_{\alpha\beta}
  +P_R 
  \left(\frac{1}{2}-\sin^2\theta_w \right)\lambda_{\alpha\beta}
  \right]\ell_\beta
  \right\}.\label{eqn:LNC_2}
\end{align}
Because matrix $\epsilon$ can be non-diagonal, the quantity inside the square brackets in (\ref{eqn:LCC_2}) is non-unitary in general. Comparing $\mathcal{L}^{NC}_2$ with (\ref{equ:L_NC_R}), it is clear that the model is now more complicated as there are two new effects ($\epsilon$ and $\lambda$) entering. In fact, many of the entries in the generalized coupling matrices (\ref{equ:g_W}) to (\ref{equ:g_H_both}) will also get modified with $M_R$ or $\epsilon$ dependent terms.
This, however, should not come as a surprise since $\nu_R$ is itself an exotic particle, very much like the doublet $\widetilde{L}_{L,R}$ (see Table~\ref{table:field_summary}).

In the light of this, we have opted to omit the additional mixing effects caused by the RH neutrino Majorana terms from our analysis. Alternatively, one may think of this as taking the limit $M_R \gg \widetilde{M}$, so that only the $\lambda$ mixing effects will be important.

\section{Constraints from $Z$ decays}\label{sec:Z_decays}

As hinted earlier, the key to any new physics contributions to the electroweak processes can be parametrized by the elements of the $\lambda$ matrix  for  they incorporate all the essential information regarding  the exotic doublets $\widetilde{L}$ (see (\ref{eqn:def_lambda}) for definition). Therefore, it is useful to  study the phenomenological constraints on the entries of $\lambda$ from precision measurements. 
In this section, we investigate the bounds coming from tree-level $Z$ decays to charged leptons: $Z \rightarrow \ell_\alpha \overline{\ell}_\beta$. 
The $\alpha=\beta$ cases will place restrictions on $\lambda_{\alpha\alpha}$'s whereas for $\alpha \neq \beta$, the off-diagonal entries can be constrained.
 
Although the limits presented here for the off-diagonal elements will not be as stringent as those obtained from other LFV interactions (see Sec.~\ref{sec:L23L}, \ref{sec:L2eg} and \ref{sec:m_to_e}), 
the constraints for the diagonal elements of $\lambda$ will be useful in the later analysis of the anomalous magnetic moments (see Sec.~\ref{sec:AMM}).

Calculating the decay rate, $\Gamma(Z\rightarrow \ell_\alpha \overline{\ell}_\alpha)$, using the standard method but with the modified couplings in (\ref{equ:L_NC_R}), we obtain (in the limit of massless final state leptons)
\begin{align}
 \Gamma(Z\rightarrow \ell_\alpha \overline{\ell}_\alpha) &=
  \frac{G_F M_Z^3}{3 \sqrt{2} \,\pi}\left(
  \left|-\frac{1}{2} + \sin^2 \theta_w\right|^2 +
  \left|\sin^2 \theta_w + 
    \left(\frac{1}{2} - \sin^2 \theta_w\right)
    \lambda_{\alpha\alpha}\right|^2
  \right), \quad \alpha = e, \mu, \tau\;, \label{eqn:Z_aa_rate}
\end{align}
where $\theta_w$ and $M_Z$ are the usual Weinberg angle and $Z$ boson mass respectively.
Inserting the decay widths and values of the constants obtained from experiments \cite{Nakamura:2010zzi}, Eq.~(\ref{eqn:Z_aa_rate}) leads to the following constraints 
for each lepton flavor $\alpha$:
\begin{align}
\lambda_{\alpha\alpha} \lesssim
 \begin{cases}
  6.1\pm 0.3\times 10^{-3}\;,\quad \alpha = e\;, \\
  7.0\pm 0.3\times 10^{-3}\;,\quad \alpha = \mu\;, \\
  8.0\pm 0.3\times 10^{-3}\;,\quad \alpha = \tau\;.
 \end{cases} \label{eqn:Laa_values}
\end{align}
Having established the constraints for $\lambda_{\alpha\alpha}$ above, one may further estimate the change to the polarization asymmetry of the $Z\rightarrow \ell\overline{\ell}$ decay due to the effects of the exotic doublets. In terms of the new generalized couplings, we have
\begin{align}
 A^{\ell}_{LR} &\equiv \frac{(g_{2L}^Z)_{11}^2-(g_{2R}^Z)_{11}^2}
                         {(g_{2L}^Z)_{11}^2+(g_{2R}^Z)_{11}^2}\;,
                         \nonumber\\
           &= \frac{1/4 - \sin^2 \theta_w -2\sin^2  \theta_w (1/2 - \sin^2  \theta_w) \lambda_{\alpha\alpha} + \order{\lambda^2}}
           {1/4 - \sin^2 \theta_w +2\sin^4 \theta_w +2\sin^2  \theta_w (1/2 - \sin^2  \theta_w) \lambda_{\alpha\alpha} + \order{\lambda^2}}
           \;,
           \label{eqn:Z_pol_asym}              
\end{align} 
where the subscript ``11'' indicates the appropriate matrix element. Note that the SM prediction for this asymmetry can be recovered from (\ref{eqn:Z_pol_asym}) by setting $\lambda_{\alpha\alpha} =0$. At the $Z$ resonance, it is well-known that the forward-backward asymmetry for the process $e^+e^- \rightarrow \ell^+ \ell^-$ is related to $A^\ell_{LR}$ simply via
\begin{equation}
 A^{0,\ell}_{FB} = \frac{3}{4}\, (A^{\ell}_{LR})^2\;.
\end{equation}
For our exotic doublet model with $\sin^2 \theta_w = 0.231$, this quantity is evaluated to about $1.6\times 10^{-2}$ if $\lambda_{\alpha\alpha} = 6.0\times 10^{-3}$ while it is about $1.7\times 10^{-2}$ if $\lambda_{\alpha\alpha} = 1.0\times 10^{-3}$. Comparing this with the experimental best fit \cite{Nakamura:2010zzi}, $1.71\pm 0.10 \times 10^{-2}$, one can see that the limits established in (\ref{eqn:Laa_values}) are more or less consistent with this precision test. The small deviation which appears between the two choices of $\lambda_{\alpha\alpha}$ demonstrated above indicates that for all calculations, the most conservative approach is to interpret (\ref{eqn:Laa_values}) as
\begin{equation}\label{eqn:Laa_values2}
 \lambda_{\alpha\alpha} \lesssim 10^{-3} \quad\text{for all } \alpha\;. 
\end{equation}

Next, we consider the case where $\alpha \neq \beta$. The decay rate is given by
\begin{align}\label{eqn:Zab_rate}
 \Gamma(Z\rightarrow \ell_\alpha \overline{\ell}_\beta) &=
 \frac{G_F M_Z^3}{3 \sqrt{2} \,\pi} 
 \left(\frac{1}{2}-\sin^2\theta_w \right)^2
 \left|\lambda_{\alpha\beta} \right|^2\;, \qquad \alpha \neq \beta \;.
\end{align}
Note that in the limit $\lambda_{\alpha\beta} \rightarrow 0$, this rate disappears in accordance with the fact that there is no flavor changing neutral currents (FCNC) at tree-level in the SM.
Writing this as a branching ratio and keeping only the leading order terms, one obtains
\begin{align}
 \BR{Z\rightarrow \ell_\alpha \overline{\ell}_\beta} &=
  \frac{\drate{Z\rightarrow \ell_\alpha \overline{\ell}_\beta}}
  {\Gamma(Z\rightarrow \ell_\sigma \overline{\ell}_\sigma)}\,\BR{Z\rightarrow \ell_\sigma \overline{\ell}_\sigma}\;,
    \nonumber
 \\
 &\simeq \frac{\left|\lambda_{\alpha\beta} \right|^2
 (4 \sin^4 \theta_w - 4\sin^2 \theta_w + 1)}
 {8 \sin^4 \theta_w - 4\sin^2 \theta_w + 1}\,\BR{Z\rightarrow \ell_\sigma \overline{\ell}_\sigma}
    \;.\label{eqn:Zab_BR}
\end{align}
From this, we can derive the following bounds for $|\lambda_{\alpha\beta}|$:\footnote{Note that the LFV branching ratios quoted in \cite{Nakamura:2010zzi} is in fact the experimental values for $\BR{Z\rightarrow \ell_\alpha \overline{\ell}_\beta} + \BR{Z\rightarrow \overline{\ell}_\alpha \ell_\beta }$. Therefore, the expression in (\ref{eqn:Zab_BR}) must be multiplied by a factor of 2 before applying the experimental numbers.}
\begin{align}
  &|\lambda_{e\mu}|< 6.6 \times 10^{-3}\;, \label{eqn:Z_offd_1}\\
 &|\lambda_{e\tau}|< 1.6 \times 10^{-2}\;,\label{eqn:Z_offd_2}\\
&|\lambda_{\mu\tau}|< 1.8 \times 10^{-2}\;.\label{eqn:Z_offd_3}
\end{align}
Because we are working in the basis where $\widetilde{M}$ is real and diagonal, $\lambda$ is necessarily hermitian. Therefore, we have $|\lambda_{\alpha\beta}|=|\lambda_{\beta\alpha}|$.
As a result, all entries of the $\lambda$ matrix can now be constrained. 

But as foreshadowed, we know from past experience that some of the strongest bounds on such new physics would come from the lepton flavor violating decays of ordinary charged leptons. So, processes like $\ell \rightarrow 3\ell'$ and $\ell \rightarrow \ell' \gamma$ are expected to yield even stronger bounds for $|\lambda_{\alpha\beta}|$ than those presented in this section. Moreover, we anticipate that the strongest limit on $|\lambda_{e\mu}|$
will come from the studies of muon-to-electron ($\mu$-$e$) conversion in atomic nuclei as it is well-known  that this process gives rise to a very strong constraint on the $\mu$-$e$-$Z$ vertex \cite{Bernabeu:1993ta}. We shall investigate these in following sections.

\section{tree-level $\ell \rightarrow 3\ell'$ decays}\label{sec:L23L}

Assuming three generations of ordinary leptons, there are only three generic types of final lepton states possible for a charged lepton decaying into three lighter ones: $\ell_\beta \ell_\beta \overline{\ell}_\beta$,
$\ell_\sigma \ell_\beta \overline{\ell}_\beta$ and
$\overline{\ell}_\sigma \ell_\beta \ell_\beta$, where $\beta\neq\sigma\neq\alpha$, with $\alpha$ denoting the flavor of the decaying lepton. In theory, the mediating particle here can  either be the gauge  boson $Z$ or the Higgs boson $H$. However, since the amplitude associated with the Higgs is suppressed by a factor of $m_\alpha^2/M_H^2$ (where $m_\alpha$ and $M_H$ denote the lepton and Higgs masses respectively), it is safe to neglect their contributions for this analysis.

So from (\ref{equ:L_NC_R}), we can write down the branching ratios:
\begin{align}
 \BR{\ell_\alpha \rightarrow \ell_\beta \ell_\beta \overline{\ell}_\beta}
    &= \frac{\drate{\ell_\alpha \rightarrow \ell_\beta \ell_\beta \overline{\ell}_\beta}}
    {\drate{\ell_\alpha \rightarrow \ell_\beta \nu_\alpha \overline{\nu}_\beta}}\,
    \BR{\ell_\alpha \rightarrow \ell_\beta \nu_\alpha \overline{\nu}_\beta}\;, \nonumber\\
    &\simeq |\lambda_{\beta\alpha}|^2 (12\sin^8\theta_w -12\sin^6\theta_w +3\sin^4\theta_w)\,
    \BR{\ell_\alpha \rightarrow \ell_\beta \nu_\alpha \overline{\nu}_\beta}\;,
    &\text{ for } \alpha = \mu, \tau\;,
    \label{eqn:BR_abbb}\\
\intertext{and for $\alpha = \tau$ only}
   \BR{\ell_\alpha \rightarrow \ell_\sigma \ell_\beta \overline{\ell}_\beta}
    &\simeq |\lambda_{\sigma\alpha}|^2 (8\sin^8\theta_w -8\sin^6\theta_w +2\sin^4\theta_w)\,
    \BR{\ell_\alpha \rightarrow \ell_\beta \nu_\alpha \overline{\nu}_\beta}\;,&
    \label{eqn:BR_asbb} 
   \\
   \BR{\ell_\alpha \rightarrow \overline{\ell}_\sigma \ell_\beta \ell_\beta}
    &\simeq 2 |\lambda_{\beta\sigma}|^2 |\lambda_{\beta\alpha}|^2
    (\sin^2\theta_w -1/2)^4 \,
    \BR{\ell_\alpha \rightarrow \ell_\beta \nu_\alpha \overline{\nu}_\beta}\;,
    &
    \label{eqn:BR_absb}
\end{align}
where we have kept only the leading order terms.

Using the data from \cite{Nakamura:2010zzi}, we can derive the following limits for the various elements of $\lambda$.
In (\ref{eqn:BR_abbb}), there are three kinematically allowed processes
($\mu \rightarrow 3e, \tau \rightarrow 3e, \tau \rightarrow 3\mu$) and one gets
\begin{align}
 &|\lambda_{e\mu}|< 4.7 \times 10^{-6}\;, \label{eqn:Lee_bound_tree}\\
 &|\lambda_{e\tau}|< 2.1 \times 10^{-3}\;,\label{eqn:Let_bound_tree}\\
&|\lambda_{\mu\tau}|< 2.0 \times 10^{-3}\;,\label{eqn:Lmt_bound_tree}
\end{align}
while (\ref{eqn:BR_asbb}) has two possibilities ($\tau \rightarrow e \mu\overline{\mu}$ and
$\tau \rightarrow \mu e\overline{e}$), yielding
\begin{align}
 &|\lambda_{e\tau}|<3.4 \times 10^{-3} \;,\\
&|\lambda_{\mu\tau}|< 2.2 \times 10^{-3}\;.
\end{align}
Finally, we have
\begin{align}
 &|\lambda_{\mu e}||\lambda_{\mu\tau}|< 3.6 \times 10^{-3}\;,\\
&|\lambda_{e\mu}||\lambda_{e\tau}|< 3.3 \times 10^{-3}\;,
\label{eqn:LemLet_bound_tree}
\end{align}
from another two possibilities ($\tau \rightarrow \overline{e}\mu\mu$ and
$\tau \rightarrow \overline{\mu}e e$) allowed by (\ref{eqn:BR_absb}). 

As expected, these LFV processes provide a stronger set of constraints than those derived in (\ref{eqn:Z_offd_1}) to (\ref{eqn:Z_offd_3}) from the previous section.

\section{radiative $\ell \rightarrow \ell'\gamma$ decays via one loop}\label{sec:L2eg}

\begin{figure}[tb]
\begin{center}
\includegraphics[width=0.25\columnwidth]{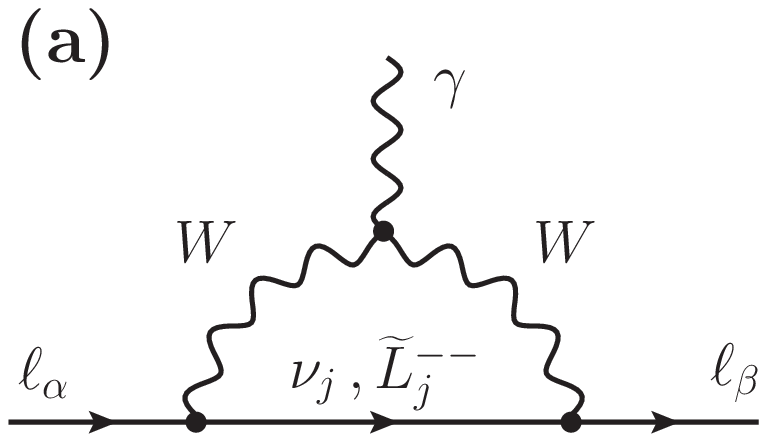}
\includegraphics[width=0.25\columnwidth]{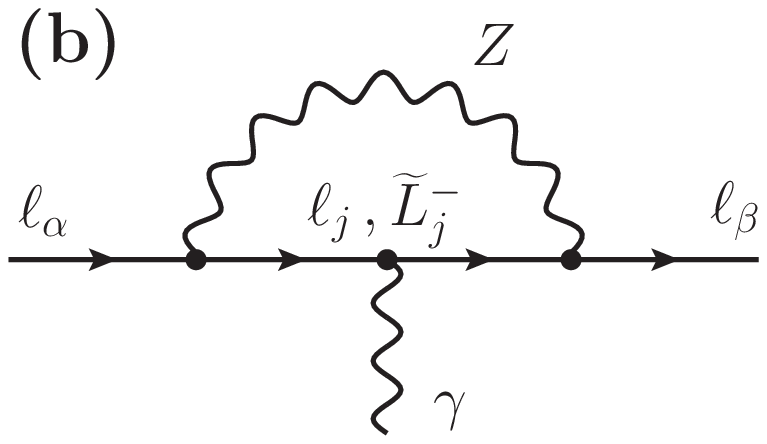}
\includegraphics[width=0.25\columnwidth]{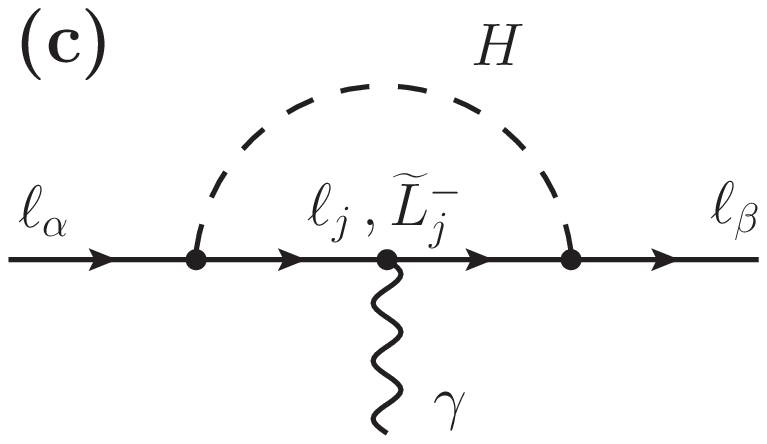}
\caption{Lowest-order diagrams that are relevant for the amplitude calculations of LFV decays ($\ell_\alpha \rightarrow \ell_\beta \gamma$) and anomalous magnetic moments of SM leptons (when $\alpha = \beta$) in the unitary gauge. Subscript $j$ denotes the flavor of the internal leptons, and is summed over in the computation. (a) The case mediated by $\nu_j$ corresponds to the usual diagram studied in standard electroweak theory, while the $\widetilde{L}^{--}_j$ diagram comes from the new interactions; (b) \& (c) are new contributing diagrams involving the $Z$ boson and physical Higgs, $H$, respectively.}
\label{fig:WZH_loop}
\end{center}
\end{figure}

It is clear that given the continual experimental effort on improving the bounds associated with LFV radiative decays of charged leptons ($\ell \rightarrow \ell'\gamma$),\footnote{See for example the review in \cite{Marciano:2008zz}.}
any new contribtutions to these interactions originating from the exotic doublets  must not be overlooked. Therefore, in this section, we calculate the effects due to having additional doublet particles $\widetilde{L}^{-}$ and $\widetilde{L}^{--}$ running inside the one-loop diagrams as depicted in Fig.~\ref{fig:WZH_loop}.

To set our notation, consider the following generic transition amplitude for $\ell_\alpha \rightarrow \ell_\beta \gamma$:
\begin{equation}\label{eqn:generic_EMDM_term}
 T(\ell_\alpha \rightarrow \ell_\beta\gamma) =
  \overline{u}_\beta \,
 (A+B\gamma_5)
  \, i \sigma_{\rho\nu}  q^\nu \varepsilon^\rho u_\alpha\;, \qquad \sigma_{\rho\nu} \equiv i \left[\gamma_\rho, \gamma_\nu\right]/2\;,
\end{equation}
where $A$ and $B$ correspond to the transition magnetic and electric dipole form factors\footnote{It is understood that $A$ and $B$ are dimensionful quantities when written in this form. Also, we have absorbed the extra factor of $i$ into $B$ which is usually factored out in the definition of the electric dipole moment term.}, while  $q^\nu$ and $\varepsilon^\rho$ denote  the photon 4-momentum and  polarization respectively.

Applying the modified coupling matrices given in (\ref{equ:g_W}) to (\ref{equ:g_H_both}), we can  easily relate the new physics parameter $\lambda_{\alpha\beta}$ to these LFV processes. When explicitly computing the diagrams in Fig.~\ref{fig:WZH_loop} in the unitary gauge,\footnote{Note that these are the only graphs we need to consider in this gauge.}
our strategy is to perform the calculations in terms of the generalized renormalizable ($R_\xi$) gauge \cite{Fujikawa:1972fe}, and subsequently, taking the limit $\xi \rightarrow \infty$ to obtain the desired results.\footnote{We have adopted the definition of $\xi$ as used in modern textbooks \cite{Cheng:1985bj, Peskin:1995ev}, which is equivalent to the parameter $1/\xi$ as appeared in \cite{Fujikawa:1972fe}.} Moreover, we will work exclusively in the $m_{\ell_j} \ll M_{W,Z,H}$ and $m_\beta \ll 1$ limits (where $m_{\ell_j}$ and $m_\beta$ represent the masses of the internal $j$-flavor and the final state SM lepton respectively), dropping any sub-leading order terms in the process.
In these limits, one also finds that amplitudes $A$ and $B$ become identical and we may simply pick out the coefficients associated with the $\overline{u}_\beta (1\pm \gamma_5)(2p\cdot \varepsilon)u_\alpha$ components  in (\ref{eqn:generic_EMDM_term}), where $p$ denotes the momentum of $\ell_\beta$, to get our final expressions.\footnote{A more detailed discussion on this procedure can be found in \cite{Chua:2010me}, as well as many standard textbooks, see e.g. \cite{Cheng:1985bj}.}

After the dust has settled and to leading order in $\lambda$, we obtain the following expressions for the amplitudes of the various one-loop contributions (superscripts and subscripts denote the type of internal leptons and bosons involved respectively):
\begin{align}
A_W^{\nu} &= \frac{i G_F m_\alpha e}{8\pi^2\sqrt{2}} 
 \sum_j \frac{m_{\nu_j}^2}{4M_W^2} \left(U_\nu\right)_{\beta j}
 \left(U_\nu^\dagger\right)_{j\alpha} \simeq 0\,, \label{eqn:LFV_amp_W_nu}
\\
A_W^{\widetilde{L}^{--}} &= \frac{i G_F m_\alpha e}{8\pi^2\sqrt{2}}
 \sum_j \frac{v^2}{2} \left(\widetilde{Y} \widetilde{M}^{-1}\right)_{\beta j} \left(\widetilde{M}^{-1} \widetilde{Y}^\dagger \right)_{j\alpha}
  \left[f_1(w_j) + f_2(w_j)\right]\,, \qquad w_j \equiv 
  \widetilde{M}_j^2/M_W^2\;,
\label{eqn:LFV_amp_W_E}
\\
A_Z^{\ell} &= \frac{i G_F m_\alpha e}{8\pi^2\sqrt{2}} \,
 \left(1- \frac{8}{3} \sin^2\theta_w +\frac{4}{3} \sin^4\theta_w
 \right) \lambda_{\beta\alpha}\;,
\label{eqn:LFV_amp_Z_l}
\\
A_Z^{\widetilde{L}^-} &= \frac{i G_F m_\alpha e}{8\pi^2\sqrt{2}}
 \sum_j \frac{v^2}{2} \left(\widetilde{Y} \widetilde{M}^{-1}\right)_{\beta j} \left(\widetilde{M}^{-1} \widetilde{Y}^\dagger \right)_{j\alpha}
  \left[f_{3A}(z_j) +f_{3B}(z_j) + 2f_4(z_j)+ 3f_5(z_j)\right]\,, \quad z_j \equiv \widetilde{M}_j^2/M_Z^2\;,
\label{eqn:LFV_amp_Z_E}
\\
\intertext{}
A_H^{\ell}&= \frac{i G_F m_\alpha e}{8\pi^2\sqrt{2}}\;
 \order{\frac{m_{\alpha}^2}{M_H^2} \ln \left[\frac{m_{\alpha}^2}{M_H^2}\right]}
 \lambda_{\beta\alpha} \simeq 0\;,
\label{eqn:LFV_amp_H_l}
\\
A_H^{\widetilde{L}^-} &= \frac{i G_F m_\alpha e}{8\pi^2\sqrt{2}}
 \sum_j \frac{v^2}{2} \left(\widetilde{Y}\right)_{\beta j} \widetilde{M}_j^{-2} \left(\widetilde{Y}^\dagger\right)_{j\alpha}
  \left[2 f_5(h_j) +f_6(h_j) \right]\,, \qquad h_j \equiv \widetilde{M}_j^2/M_H^2\;,
\label{eqn:LFV_amp_H_E}
\end{align}
with
\begin{align}
 f_1(x) &= \frac{-10+43x -78x^2 +49x^3 -4x^4 -18x^3 \ln x}{12 (x-1)^4} \;,
 \label{eqn:f_1}
\\
 f_2(x) &= \frac{4 -15x +12x^2 -x^3 -6x^2 \ln x}{2 (x-1)^3} \;,
 \label{eqn:f_2}
\\
 f_{3A}(x) &= \frac{4-9x +5x^3- 6x(2x-1)\ln x}{12 (x-1)^4} \;,
 \label{eqn:f_3A}
\\
 f_{3B}(x) &= \frac{2(1-x^2+2x\ln x)}{(x-1)^3} \;,
 \label{eqn:f_3B} 
\\
%
 f_4(x) &= \frac{-11x +18x^2 - 9x^3 +2x^4 -6x \ln x}{48 (x-1)^4} \;,
 \label{eqn:f_4}
\\
 f_5(x) &= \frac{-3x +4x^2 -x^3  -2x \ln x}{8 (x-1)^3} \;,
 \label{eqn:f_5}
\\
 f_6(x) &= \frac{-2x -3x^2 +6x^3- x^4 -6x^2 \ln x}{24 (x-1)^4} \;.
 \label{eqn:f_6}
\end{align}
In the above, $m_{\nu_j}$, $m_{\alpha}$ and $\widetilde{M}_{j}$ denote, respectively, the masses of the $j$-flavor neutrino, the decaying SM lepton $\ell_\alpha$ and the exotic $\widetilde{L}$ particle. Note that (\ref{eqn:LFV_amp_W_nu}) is nothing but the contribution due to the SM electroweak interactions when neutrinos carry a nonzero mass. It is well-known  that the size of this is negligible \cite{Cheng:1985bj, mu2eg_SM, Bilenky:1987ty} as it receives a $m_{\nu_j}^2/M_W^2$ suppression.

With these amplitudes and the general formula for the total decay rate,
\begin{equation}
 \drate{\ell_\alpha \rightarrow \ell_\beta \gamma} = \frac{m_\alpha^3}{4\pi}
 \left|A_W^{\nu}+ A_W^{\widetilde{L}^{--}}
  +A_Z^{\ell}+A_Z^{\widetilde{L}^{-}}+A_H^{\ell}
  +A_H^{\widetilde{L}^{-}}\right|^2 \;,
 \label{eqn:mu2g_rate}
\end{equation}
we obtain the branching ratio (after dropping $A_W^{\nu},A_H^{\ell} \simeq 0$)
\begin{align}
 \BR{\ell_\alpha \rightarrow \ell_\beta \gamma} &=
  \frac{3\alpha_e}{2\pi} \left| 
 \left(1- \frac{8}{3} \sin^2\theta_w +\frac{4}{3} \sin^4\theta_w
 \right)\lambda_{\beta\alpha}
 + \sum_j \frac{v^2}{2} \left(\widetilde{Y}\right)_{\beta j} \widetilde{M}_j^{-2} \left(\widetilde{Y}^\dagger\right)_{j\alpha}
  \left[\phantom{\frac{}{}}f_1(w_j) + f_2(w_j) 
   \right. 
 \right.
     \nonumber\\
 &\qquad\qquad\qquad
 \left.\left. + f_{3A}(z_j)  + f_{3B}(z_j)
 +  f_4(z_j)+ 3f_5(z_j)+2f_5(h_j) +f_6(h_j) \phantom{\frac{}{}}\right]
 \phantom{\sum_j\hspace{-5mm}}\right|^2
 \BR{\ell_\alpha \rightarrow \ell_\beta \nu_\alpha \overline{\nu}_\beta}\;,
\label{eqn:mu2g_BR}
\end{align}
where $\alpha_e$ is the fine-structure constant.
Taking $\widetilde{M}_{j} \simeq \order{100}$~GeV (this is around the global lower bound for heavy charged leptons from current experiments \cite{Nakamura:2010zzi})\footnote{Note that 
the resulting bounds will become less stringent
as we increase the value for $\widetilde{M}_j$. Also, we would like to remind the reader that the reason we need to specify a size for $\widetilde{M}_j$ here is solely for the evaluation of the loop functions (\ref{eqn:f_1}) to (\ref{eqn:f_6}). We have checked that the value for these loop functions would only change by a small amount even when we take the very large $\widetilde{M}_j$ limit.} 
for all $j$, and assuming the Higgs mass, $M_H$, is about 114~GeV \cite{Barate:2003sz}, the experimental limits \cite{Nakamura:2010zzi} on \BR{\mu\rightarrow e\gamma}, \BR{\tau\rightarrow e\gamma} and \BR{\tau\rightarrow \mu\gamma} then lead to\footnote{If we compare this set of limits with the corresponding ones in \cite{Chua:2010me}, we notice that these bounds are somewhat stronger. The reason for this comes from the fact that \cite{Chua:2010me} omitted the additional $W$-mediated graph where the SM leptons couple to an internal \emph{doubly} charged exotic triplet. As a result, the small accidental cancellations in the numerics (as happened in these numbers here) did not happen there.}
\begin{align}
 &|\lambda_{e\mu}|< 2.4 \times 10^{-5}\;, \label{eqn:Lee_bound_loop}\\
 &|\lambda_{e\tau}|< 3.0 \times 10^{-3}\;, \label{eqn:Let_bound_loop}\\
&|\lambda_{\mu\tau}|< 3.5 \times 10^{-3}\; \label{eqn:Lmt_bound_loop}.
\end{align}
Although these bounds are weaker than those displayed in (\ref{eqn:Lee_bound_tree}) to (\ref{eqn:Lmt_bound_tree}), the derived expressions above will be very useful when the expected improvement in the experimental bounds are realized in the near future (besides the upper limits, (\ref{eqn:Let_bound_loop}) and (\ref{eqn:Lmt_bound_loop}), are only marginally bigger than their counterparts). Currently, the MEG experiment \cite{exp_MEG} located at Paul Scherrer Institute (PSI) is planning to reach a sensitivity of \order{10^{-13}} for the $\mu \rightarrow e \gamma$ branching ratio, which is a significant improvement compare to the current limit of \BR{\mu\rightarrow e\gamma} $< 1.2 \times 10^{-11}$ \cite{Brooks:1999pu}. In addition, the Super KEKB project \cite{exp_SuperB} will provide an excellent platform for investigating LFV $\tau$ decays at an unprecedented precision. As a result, the bounds on $\tau \rightarrow e\gamma$ and $\tau\rightarrow \mu\gamma$ are also expected to tighten, providing a stronger constraint for all off-diagonal couplings $\lambda_{\alpha\beta}$ than presented here.

\section{$\mu$-$e$ conversion in atomic nuclei}\label{sec:m_to_e}

Muon-to-electron conversion in muonic atoms provides another excellent testing field for tree-level FCNC. This is because coherent contribution of all nucleons in the nucleus can enhance the experimental signals, and hence the $\mu$-$e$-$Z$ vertex may be probed at great precision. Given that this is the same vertex as appeared in  the loop graphs in Fig~\ref{fig:WZH_loop}b, the test for $\mu$-$e$ conversion, therefore, plays a complementary role to the investigation of $\mu\rightarrow e\gamma$ in the probe for physics beyond the SM as the two processes are induced differently.

In what follows, we shall assume that the only contribution to the $\mu$-$e$ conversion rate in our setup comes from exchanges with the $Z$ bosons. This approximation is sensible because the cases mediated by the photon and the Higgs are suppressed by loop effects and $M_H^{-1}$ respectively. So, assuming only SM interactions operate in the quark sector, we obtain the following effective interaction Lagrangian for the $\mu$-$e$ transition (after integrating out $M_Z$):
\begin{align}
 \mathcal{L}^\text{eff}_{\mu\rightarrow e}
  &= \sqrt{2} \, G_F \overline{\ell}_e \gamma^\nu (k_V - k_A \gamma_5) \ell_\mu
  \left[
  \overline{q}_u \gamma_\nu (v_u + a_u \gamma_5) q_u
  +
  \overline{q}_d \gamma_\nu (v_d + a_d \gamma_5) q_d
  \right] \;, \label{eqn:m2e_L_eff}
\end{align}
where $q_{u,d}$ denotes the $u, d$-quark field while
\begin{eqnarray}
 k_V = -k_A = \left(\frac{1}{2}- \sin^2 \theta_w\right)\lambda_{e\mu}\;, \quad
 a_u = -a_d = -\frac{1}{2}\;,\quad
 v_u = \frac{1}{2}-\frac{4}{3} \sin^2 \theta_w \;, \quad
 v_d = -\frac{1}{2}+\frac{2}{3} \sin^2 \theta_w \;.
\end{eqnarray}

Appealing to the general result obtained from FCNC analysis with massive gauge bosons in \cite{Bernabeu:1993ta}, the branching ratio for $\mu$-$e$ conversion in nuclei (for nuclei with less than about 100 nucleons) is found to be
\begin{equation}\label{eqn:m2e_BR_formula}
  B_{\mu\rightarrow e} \simeq
  \frac{G_F^2 \,\alpha_e^3 \,m_\mu^3\, p_e'\, E_e'}{\pi^2\,\Gamma_\text{cap}^\mathcal{A}}
  \left|F(q'^2)\right|^2
  \left(k_V^2+k_A^2\right) \frac{\mathcal{Z}_\text{eff}^4\, \hat{Q}^2}{\mathcal{Z}} \;,
\end{equation}
where $p_e'$ ($E_e'$) is the momentum (energy) of the electron, $\Gamma_\text{cap}^\mathcal{A}$ represents the total nuclear muon capture rate for element $\mathcal{A}$, and
$\mathcal{Z}$ ($\mathcal{Z}_\text{eff}$) is the (effective) atomic number of the element under investigation. In (\ref{eqn:m2e_BR_formula}),
$F(q'^2)$ is the nuclear form factor which may be measured from electron scattering experiments \cite{11of_ref22} while %
\begin{equation}
 \hat{Q} = (2\mathcal{Z} + \mathcal{N}) \, v_u + (\mathcal{Z} + 2\mathcal{N})\, v_d \;,
\end{equation}
with $\mathcal{N}$ denoting the number of neutrons in the nuclei.

Given that one of the best upper limit on the $\mu$-$e$ conversion branching ratio is obtained from measurements with titanium-48 ($^{48}_{22}\text{Ti}$) in the SINDRUM II experiments \cite{Dohmen:1993mp}:
\begin{equation}\label{eqn:m2e_BR_Ti}
 B_{\mu\rightarrow e}^\text{exp} \equiv
 \frac{\drate{\mu^-\,\text{Ti} \rightarrow e^-\,\text{Ti}}}{\Gamma_\text{cap}^\text{Ti}}
 < 4.3 \times 10^{-12}\;,
\end{equation}
we shall use the parameters for element $^{48}_{22}\text{Ti}$ in (\ref{eqn:m2e_BR_formula}) to deduce our bound.\footnote{Although the value quoted in the experiments with gold (Au):
$\drate{\mu^-\text{Au} \rightarrow e^-\text{Au}} / \Gamma_\text{cap}^\text{Au} <
7\times 10^{-13}$ \cite{Bertl:2006up} is smaller than the one in (\ref{eqn:m2e_BR_Ti}), theoretical calculations \cite{Kitano:2002mt} have shown that for very heavy elements (atomic number $\mathcal{Z} \gtrsim 60$) like Au, the $\mu$-$e$ conversion rate is actually suppressed. Therefore, this does not necessarily indicate a better bound on the rate, especially when the estimation of the nuclear matrix element for such heavy nuclei can carry large uncertainties.}
Following the approximation as applied in \cite{Bernabeu:1993ta}, we take $p_e'\simeq E_e' \simeq m_\mu$, and $F(q'^2 \simeq -m_\mu^2) \simeq 0.54$. In addition, we have $\mathcal{Z}_\text{eff} \simeq 17.6$ for $^{48}_{22}\text{Ti}$ \cite{Zeff} and $\Gamma_\text{cap}^\mathcal{A} \equiv \Gamma_\text{cap}^\text{Ti} \simeq 2.59\times 10^6$~s$^{-1}$ \cite{Suzuki:1987jf}. Hence, (\ref{eqn:m2e_BR_formula}) and (\ref{eqn:m2e_BR_Ti}) combine to give
\begin{equation}\label{eqn:m2e_Lem_bound}
 |\lambda_{e\mu}| < 1.9 \times 10^{-6}\;.
\end{equation}
As hinted earlier, the bound displayed in (\ref{eqn:m2e_Lem_bound}) is indeed the most stringent one on $|\lambda_{e\mu}|$. Moreover, given that new $\mu$-$e$ experiments are being planned, respectively, at J-PARC and Fermilab by the COMET (and PRISM/PRIME) \cite{exp_COMET} and Mu2e \cite{exp_mu2e} collaborations, this bound is expected to be further strengthened in the near future.

\section{Global fit on the elements of $\lambda$ and some consequences}\label{sec:global_fit}

In this section, we bring together all the results obtained thus far and perform a global analysis on the
elements of the $\lambda$ matrix, which are key to determining the new physics effects from the doublet leptons, $\widetilde{L}$. For convenience, a summary of all constraints derived in the previous four sections are collected in Table~\ref{table:all_results}.

\begin{table}[ht]
\begin{center}
\begin{tabular}{|c|c|c|c|}
\hline
parameter(s) & process & limit on BR & constraint on $\lambda$'s\\
\hline
$\lambda_{ee}$ & $Z\rightarrow e^- e^+$ &$3.363\pm 0.004 \times 10^{-2}$&  $\; \; $\\
$\lambda_{\mu\mu}$ & $Z\rightarrow \mu^- \mu^+$ &$3.366\pm 0.007 \times 10^{-2}$&  $\; \lesssim 10^{-3}\; $\\
$\lambda_{\tau\tau}$ & $Z\rightarrow \tau^- \tau^+$ &$3.369\pm 0.008 \times 10^{-2}$&  $\; \; $\\
\hline
$|\lambda_{e\mu}|$ &
\begin{tabular}{c}
 $Z\rightarrow e^{\pm}\mu^{\mp}$\\
 $\mu^- \rightarrow e^- e^- e^+$\\
 $\mu\rightarrow e \gamma$\\
 $\mu$-$e$ conversion\\
\end{tabular}
&
\begin{tabular}{c}
 $<1.7 \times 10^{-6}$\\
 $<1.0 \times 10^{-12}$\\
 $<1.2 \times 10^{-11}$\\
 $<4.3 \times 10^{-12}$ (Ti)\\
\end{tabular}
&
\begin{tabular}{c}
 $< 6.6 \times 10^{-3}$  \\
 $< 4.7 \times 10^{-6}$  \\
 $< 2.4 \times 10^{-5}$  \\
 $< 1.9 \times 10^{-6}$  \\
\end{tabular} \\
\hline
$|\lambda_{e\tau}|$ &
\begin{tabular}{c}
 $Z\rightarrow e^{\pm}\tau^{\mp}$\\
 $\tau^- \rightarrow e^- e^- e^+$\\
 $\tau^- \rightarrow e^- \mu^- \mu^+$\\
 $\tau\rightarrow e \gamma$\\
\end{tabular}
&
\begin{tabular}{c}
 $<9.8 \times 10^{-6}$\\
 $<3.6 \times 10^{-8}$\\
 $<3.7 \times 10^{-8}$\\
 $<3.3 \times 10^{-8}$\\
\end{tabular}
&
\begin{tabular}{c}
 $< 1.6 \times 10^{-2}$  \\
 $< 2.1 \times 10^{-3}$  \\
 $< 3.4 \times 10^{-3}$  \\
 $< 3.0 \times 10^{-3}$  \\
\end{tabular}
\\
\hline
$|\lambda_{\mu\tau}|$ &
\begin{tabular}{c}
 $Z\rightarrow \mu^{\pm}\tau^{\mp}$\\
 $\tau^- \rightarrow \mu^- \mu^- \mu^+$\\
 $\tau^- \rightarrow \mu^- e^- e^+$\\
 $\tau\rightarrow \mu \gamma$\\
\end{tabular}
&
\begin{tabular}{c}
 $<1.2 \times 10^{-5}$\\
 $<3.2 \times 10^{-8}$\\
 $<2.7 \times 10^{-8}$\\
 $<4.4 \times 10^{-8}$\\
\end{tabular}
&
\begin{tabular}{c}
 $< 1.8 \times 10^{-2}$  \\
 $< 2.0 \times 10^{-3}$  \\
 $< 2.2 \times 10^{-3}$  \\
 $< 3.5 \times 10^{-3}$  \\
\end{tabular}
\\
\hline
$|\lambda_{\mu e}||\lambda_{\mu\tau}|$ & $\tau\rightarrow e^+ \mu^-\mu^-$ &$<2.3 \times 10^{-8}$&  $< 3.6 \times 10^{-3} $\\
$|\lambda_{e\mu}||\lambda_{e\tau}|$ & $\tau\rightarrow \mu^+ e^- e^-$ &$<2.0 \times 10^{-8}$&  $< 3.3 \times 10^{-3}$\\
\hline
\end{tabular}
\caption{A collection of all constraints on the elements of $\lambda \equiv v^2 \widetilde{Y} \widetilde{M}^{-2} \widetilde{Y}^\dagger/2$ from processes studied in the previous four sections.}\label{table:all_results}
\end{center}
\end{table}

Studying the results listed in Table~\ref{table:all_results} and recalling that $|\lambda_{\alpha\beta}| =|\lambda_{\beta\alpha}|$, it is not difficult to obtain the following overall fit for the elements of $\lambda$:
\begin{equation}\label{eqn:global_constraint}
 \ththMat{|\lambda_{ee}|}{|\lambda_{e\mu}|}{|\lambda_{e\tau}|}
 {|\lambda_{\mu e}|}{|\lambda_{\mu \mu}|}{|\lambda_{\mu \tau}|}
 {|\lambda_{\tau e}|}{|\lambda_{\tau \mu}|}{|\lambda_{\tau \tau}|}
 \lesssim
 \ththMat
 { 10^{-3}}{ 1.9 \times 10^{-6}}{ 2.1 \times 10^{-3}}
 { 1.9 \times 10^{-6}}{10^{-3}}{ 2.0 \times 10^{-3}}
 { 2.1 \times 10^{-3}}{ 2.0 \times 10^{-3}}{10^{-3}}\;.
\end{equation}
Result (\ref{eqn:global_constraint}) is one of our major results in this work. Using the definition, $\lambda \equiv v^2 \widetilde{Y} \widetilde{M}^{-2} \widetilde{Y}^\dagger/2$, and taking  $\widetilde{M} \simeq \order{100}$~GeV (for all flavors), a rough estimate of the size for the new couplings $|\widetilde{Y}_{ij}|$ may be obtained:
\begin{equation}
 \left|\widetilde{Y}_{ij}\right| \lesssim \order{10^{-2}}\text{ to } \order{10^{-3}}\;, \qquad
 \text{for all } i,j\;.
\end{equation}
If the exotic particle mass $\widetilde{M}$ is heavier than the value used above, the upper limit for $|\widetilde{Y}_{ij}|$ will be increased.

Furthermore, if we assume that the new flavor changing physics due to the presence of these exotic $\widetilde{L}$'s are the \emph{only} source of LFV, one may derive model-specific bounds on the various processes predicted by this model from studying the ratio between the different branching ratios:
\begin{align}
 \BR{Z\rightarrow e^{\pm}\mu^{\mp}} &\simeq 4.8\times 10^{-1} \;\BR{\mu\text{-}e \,\text{conversion in Ti}}\;, \nonumber\\
 \BR{\mu^- \rightarrow e^- e^- e^+} &\simeq 3.9\times 10^{-2} \;\BR{\mu\text{-}e \,\text{conversion in Ti}}\;, \nonumber\\
 \BR{\mu \rightarrow e \gamma} &\simeq 1.9\times 10^{-2} \;\BR{\mu\text{-}e \,\text{conversion in Ti}}\;,
\label{eqn:BR_relate_em}
\end{align}
for the processes involving $|\lambda_{e\mu}|^2$. Whereas for $|\lambda_{e\tau}|^2$ and  $|\lambda_{\mu\tau}|^2$, one may write
\begin{align}
 \BR{Z\rightarrow e^{\pm}\tau^{\mp}} &\simeq 7.0\times 10^{1} \; \BR{\tau^- \rightarrow e^- e^- e^+}\;, \nonumber\\
  \BR{\tau^- \rightarrow e^- \mu^- \mu^+} &\simeq 6.5\times 10^{-1} \; \BR{\tau^- \rightarrow e^- e^- e^+}\;, \nonumber\\
 \BR{\tau \rightarrow e \gamma} &\simeq 4.6\times 10^{-1} \; \BR{\tau^- \rightarrow e^- e^- e^+}\;,
\label{eqn:BR_relate_et}
\intertext{and}
 \BR{Z\rightarrow \mu^{\pm}\tau^{\mp}} &\simeq 7.1\times 10^{1} \; \BR{\tau^- \rightarrow \mu^- \mu^- \mu^+}\;, \nonumber\\
  \BR{\tau^- \rightarrow \mu^- e^- e^+} &\simeq 6.9\times 10^{-1} \; \BR{\tau^- \rightarrow \mu^- \mu^- \mu^+}\;, \nonumber\\
 \BR{\tau \rightarrow \mu \gamma} &\simeq 4.6\times 10^{-1} \; \BR{\tau^- \rightarrow \mu^- \mu^- \mu^+}\;,
\label{eqn:BR_relate_mt}
\end{align}
respectively. Whenever required in the above, we have used $\widetilde{M} \simeq \order{100}$~GeV \cite{Nakamura:2010zzi}.\footnote{We have checked that taking a larger value for $\widetilde{M}$ would only change the numerical results by a small amount.}
The main point is that we can translate these into model-specific bounds for certain LFV processes which may be used to falsify this theory.
For instance,  applying the experimental limits on the right-hand side of (\ref{eqn:BR_relate_em}) implies that this model demands $\BR{\mu \rightarrow e \gamma} \lesssim 8.2 \times 10^{-14}$. Observe that this is a couple of orders stronger than the limit set by current experiments. As a result, a future detection of this LFV process above this rate will invalidate the predictions of this minimal extension to the SM, and point to the existence of other new physics in the lepton sector.
Similar conclusions may also be drawn from other processes displayed above.

\section{Contribution to lepton anomalous magnetic moment}\label{sec:AMM}

While in Dirac theory the gyromagnetic ratio of a spin-1/2 particle is predicted to have a value of $\tilde{g}^\text{dirac} = 2$, it is well-known that quantum field theory gives a correction to this number via loop effects.
The deviation from the Dirac result of 2  is usually parameterized by the dimensionless quantity ($\alpha$ denotes the flavor)
\begin{equation} \label{eqn:g-2_defn}
 a_\alpha \equiv \frac{\tilde{g}_\alpha - 2}{2}\;,\quad \text{where $\tilde{g}_\alpha$ is the actual value of the gyromagnetic ratio,}
\end{equation}
known as the \emph{anomalous magnetic moment}. It is related to the lepton magnetic dipole moment $\vec{\mu}_\alpha = -e(1+a_\alpha)/2m_\alpha \,\vec s$, where $\vec s$ is the unit spin vector. In terms of the parameters from quantum field theory, $a_\alpha \equiv F_2(q^2=0)$, when the form factor expansion for a general lepton-photon amplitude is written as
\begin{equation}\label{eqn:form_factors_expand}
 T(\ell_\alpha \rightarrow \ell_\alpha'\gamma) =
 -i e\, \overline{u}_\alpha'
 \left[ F_1(q^2) \gamma_\rho
  +\frac{F_2(q^2)}{2m_\alpha}\,i\sigma_{\rho\nu} \,q^\nu
  +\frac{F_3(q^2)}{2m_\alpha}\, \sigma_{\rho\nu} \gamma_5 \,q^\nu +\cdots
 \right]
  \varepsilon^\rho u_\alpha
   \;, \quad e>0\;,
\end{equation}
where $q^\nu$ is again the photon momentum (see (\ref{eqn:generic_EMDM_term}) for notations).\footnote{Note that the lepton electric dipole moment is proportional to $F_3(q^2=0)$.} Therefore, the precise contribution to $a_\alpha$ from the SM (and indeed any other theories) can be calculated by considering all the relevant loop diagrams for the $F_2(0)$-term.

While the anomalous magnetic moment for the electron, muon and tauon can all be very important in their own rights, given the present experimental and theoretical development, $a_\mu$ is the most interesting observable to examine. This is because when combining the fact that significant contributions to the overall predicted $a_\mu$ value come from every major sector (QED, electroweak, hadronic) of the SM \cite{g-2_review1, mg-2_SM} with the ability to experimentally measure $a_\mu$ to extremely high accuracy \cite{mg-2_exp1, mg-2_exp2}, the SM as a whole can be scrutinized, and any discrepancies between theory and experiment would be a strong indication of new physics. On the other hand, although $a_e$ have been measured to extraordinary precision (hence providing a very stringent test on QED and the value of the fine-structure constant~$\alpha_e$ \cite{eg-2_exp, eg-2_fineS}), its low sensitivity to the contributions from strong and electroweak processes means that any hypothetical modifications to these sectors (due to new physics) would not be easily detectable. As far as $a_\tau$ is concerned, even though its much heavier mass would in theory imply better sensitivity to any new physics than $a_\mu$, its usefulness has been limited by the relatively poor experimental bounds. In fact, the best current limits set by the DELPHI experiments \cite{DELPHI} are still too coarse to even check the first significant figure of $a_\tau$ from theoretical calculations.

Currently, the experimental values for $a_e$~\cite{eg-2_exp}, $a_\mu$~\cite{mg-2_exp2} and $a_\tau$~\cite{DELPHI} are given by
\begin{align}
 a_e^\text{Exp} &= 115~965~218~073(28)\times 10^{-14}\;, \label{eqn:a_e_exp}
 \\
 a_\mu^\text{Exp} &= 116~592~089(63)\times 10^{-11}\;, \label{eqn:a_mu_exp}
 \\
 a_\tau^\text{Exp} &=
 \begin{cases}
 \; < 1.3 \times 10^{-2} \;,\\
 \; > -5.2 \times 10^{-2} \;.
 \end{cases}
 \label{eqn:a_tau_exp}
\end{align}
Focusing on the muon case, one finds that the discrepancy between experiment and the SM estimate is about 4.0$\sigma$ \cite{mg-2_SM}:\footnote{Recently, this discrepancy was re-evaluated by the group in \cite{Hagiwara:2011af} and found to be about 3.3$\sigma$ only: $\Delta a_\mu = 261(80)\times 10^{-11}$.
}
\begin{equation} \label{eqn:mg-2_diff}
 \Delta a_\mu = a_\mu^\text{Exp} - a_\mu^\text{SM} = 316(79)\times 10^{-11}\;.
\end{equation}
If this difference is real (rather than caused by incorrect leading-order hadronic approximation\footnote{Although this possibility is not completely ruled out, shifting the hadronic cross-section to bridge this gap will naturally increase the tension with the lower bound on the Higgs mass, both from LEP \cite{Barate:2003sz}
and the SM vacuum stability requirement \cite{mg-2_SM}.}), then there must be some new physics at play. In the following, we investigate whether the presence of the exotic doublets can affect this quantity in a significant way.

Calculating the anomalous magnetic moment using the modified electroweak couplings of (\ref{equ:g_W}) to (\ref{equ:g_H_both}) is in fact analogous to the computation for LFV $\ell_\alpha\rightarrow\ell_\beta\gamma$ done in Sec.~\ref{sec:L2eg}. The only differences are that $\alpha=\beta$ here and we do not take the zero mass limit for the final state lepton. Otherwise, the three main types of one-loop diagrams we need to consider are as depicted in Fig.~\ref{fig:WZH_loop}.
Working in the unitary gauge again, and noting that for the magnetic moment, the part associated with $\gamma_5$ in the general amplitude
\begin{equation}
 T(\ell_\alpha \rightarrow \ell_\alpha'\gamma) =
  \overline{u}_\alpha' \,
  (C+D\gamma_5)
  \,i \sigma_{\rho\nu}  q^\nu \varepsilon^\rho u_\alpha\;, \label{eqn:gen_mag_moment}
\end{equation}
is not needed. Hence, in the computation, we pick out the terms that are proportional to $\overline{u}_\alpha' (2p\cdot \varepsilon)u_\alpha$, where $p$ is again the momentum of the incoming $\ell_\alpha$.
Employing a similar notation system as before, the amplitudes of the one-loop diagrams from Fig.~\ref{fig:WZH_loop} for the case $\alpha = \beta$ are given by (to leading order)
\begin{align}
 C_W^{\nu} &= \frac{i G_F m_\alpha e}{8\pi^2\sqrt{2}}
  \left(-\frac{5}{3}\right), \label{eqn:AMM_W_nu}
 \\
 C_W^{\widetilde{L}^{--}} &= \frac{i G_F m_\alpha e}{8\pi^2\sqrt{2}}
  \sum_j \frac{v^2}{2} \left(\widetilde{Y} \widetilde{M}^{-1}\right)_{\alpha j} \left(\widetilde{M}^{-1} \widetilde{Y}^\dagger \right)_{j\alpha}
  \left[f_7(w_j) + 3 f_8(w_j) + f_9(w_j) -1 \right], \quad w_j \equiv \widetilde{M}_j^2/M_W^2\;,
 \label{eqn:AMM_W_E}
 \\
%
  C_Z^{\ell} &= \frac{i G_F m_\alpha e}{8\pi^2\sqrt{2}}
  \left(
  \frac{2}{3}(
  1
  +2 \sin^2\theta_w   
  -4 \sin^4\theta_w
  )
  +\frac{2\lambda_{\alpha\alpha}}{3} (4\sin^4\theta_w -8\sin^2\theta_w+3)
  \right),
 \label{eqn:AMM_Z_l}
 \\
   C_Z^{\widetilde{L}^-} &= \frac{i G_F m_\alpha e}{8\pi^2\sqrt{2}}
  \sum_j \frac{v^2}{2} \left(\widetilde{Y} \widetilde{M}^{-1}\right)_{\alpha j} \left(\widetilde{M}^{-1} \widetilde{Y}^\dagger \right)_{j\alpha}
  \left[2f_{3A}(z_j)+2f_{3B}(z_j) +  f_4(z_j) + 2f_5(z_j) \right], \quad z_j \equiv \widetilde{M}_j^2/M_Z^2\;,
 \label{eqn:AMM_Z_E}
 \\
   C_H^{\ell} &= \frac{i G_F m_\alpha e}{8\pi^2\sqrt{2}}\, (1-4\lambda_{\alpha\alpha})\;
   \order{m_{\alpha}^2/M_H^2} \simeq 0\;,
   \label{eqn:AMM_H_l}
   \\
 C_H^{\widetilde{L}^-} &= \frac{i G_F m_\alpha e}{8\pi^2\sqrt{2}}
 \sum_j \frac{v^2}{2} \,\widetilde{Y}_{\alpha j} \widetilde{M}^{-2}_j \widetilde{Y}^\dagger_{j\alpha}
  \left[4 f_5(h_j)
   +2 f_6(h_j)\right]
  \,, \quad h_j \equiv \widetilde{M}_j^2/M_H^2\;,
\label{eqn:AMM_H_E}
\end{align}
where $f_{3A}(x)$ to $f_{6}(x)$ are given in (\ref{eqn:f_3A}) to (\ref{eqn:f_6}) and
\begin{align}
 f_7(x) &= \frac{7-33x +57x^2 -31x^3 +6x^2(3x-1) \ln x}{6 (x-1)^4} \;,
 \label{eqn:f_7}
\\
 f_8(x) &= \frac{1 -4x +3x^2 - 2 x^2 \ln x}{(x-1)^3} \;,
 \label{eqn:f_8}
\\
 f_9(x) &= \frac{3 -10x +21x^2 -18x^3+ 4x^4 +6x^2\ln x}{6 (x-1)^4} \;.
 \label{eqn:f_9}
\end{align}
Comparing (\ref{eqn:gen_mag_moment}) with the form factor expansion of (\ref{eqn:form_factors_expand}), the anomalous magnetic moment can be written in terms of the amplitudes computed above:
\begin{equation}\label{eqn:amp_FF_rel}
 a_\alpha \equiv F_2(0) = -\frac{2m_\alpha}{i e} \left(
 C_{W}^{\nu}+C_{W}^{\widetilde{L}^{--}}
 +C_{Z}^{\ell}+C_{Z}^{\widetilde{L}^-}+C_{H}^{\ell}+C_{H}^{\widetilde{L}^-}
 \right)\;.
\end{equation}
Note that the result given in (\ref{eqn:amp_FF_rel}) contains the usual SM electroweak component of $a_\alpha$, as well as
the contribution induced by the new physics. Examining our results, we see that the terms which are \emph{not} proportional to $\lambda_{\alpha\alpha}$ in (\ref{eqn:AMM_W_nu}) and (\ref{eqn:AMM_Z_l}) sum up to give the usual prediction of the anomalous magnetic moment from the SM \cite{Fujikawa:1972fe}.
Removing this component from (\ref{eqn:amp_FF_rel}) and using the values for $\lambda_{\alpha\alpha}$ given in Table~\ref{table:all_results}, we obtain the following estimate for the anomalous magnetic moments coming from the new physics associated with the exotic $\widetilde{L}$ particles:
\begin{align}
 \Delta a_e^{\widetilde{L}} &\simeq 6.6\times 10^{-16}\;,
  \label{eqn:a_e_result}\\
 \Delta a_\mu^{\widetilde{L}} &\simeq 3.2\times 10^{-11}\;,
  \label{eqn:a_m_result}\\
 \Delta a_\tau^{\widetilde{L}} &\simeq 1.1\times 10^{-8}\;,
  \label{eqn:a_t_result}
\end{align}
where we have again assumed $\widetilde{M}_j\simeq \order{100}$~GeV and $M_H \simeq 114$~GeV. Looking at the results (\ref{eqn:a_e_result}) to (\ref{eqn:a_t_result}), we see that these contributions are at least one order of magnitude less than the experimental errors given for the quantities listed in (\ref{eqn:a_e_exp}) to (\ref{eqn:a_tau_exp}). Therefore, these exotic $\widetilde{L}$'s cannot help to explain the muon $g-2$ anomaly nor can their effects be easily distinguishable from the SM components in these experiments.

\section{Comments on collider signatures}\label{sec:LHC}

If the exotic doublets $\widetilde{L}$ do exist, it would give rise to collider signals which may be detectable at the LHC. Given that the exotic mass $\widetilde{M}$ is not too massive and assuming some favorable conditions are met, such signatures can be quite distinctive as pointed out recently in \cite{DelNobile:2009st}. In this section, we summarize some of the key features one can expect from the presence of these exotic particles.

To make the situation more clear-cut, let suppose Yukawa couplings $\widetilde{Y}$ are small enough such that $\widetilde{L}$ production is predominantly mediated by SM gauge interactions. The relevant production mechanisms at the partonic level are then given by:
\begin{equation}
 q\overline{q} \rightarrow  \gamma, Z \rightarrow 
 \widetilde{L}^-  \widetilde{L}^+\;,\;
 \widetilde{L}^{--}  \widetilde{L}^{++}
 \quad ;\quad
 u\overline{d} \rightarrow W^+ \rightarrow 
 \widetilde{L}^{-}  \widetilde{L}^{++}
 \;.
\end{equation} 
With the planned energy of $\sqrt{s}=14$~TeV and luminosity at $300 \text{ fb}^{-1}\text{yr}^{-1}$ at the LHC, it is estimated that the pair production cross-section from $pp$ collisions ranges from 
$\sigma(pp\rightarrow \widetilde{L}\overline{\widetilde{L}}) \simeq$ a few~fb for $\widetilde{M}\simeq 500$~GeV to just under $10^3$~fb for $\widetilde{M} \simeq$~200 GeV \cite{DelNobile:2009st} (assuming no cuts are made).
Once produced, the singly and doubly charged exotic leptons may decay into SM particles via the interaction terms depicted in (\ref{equ:L_CC}) to (\ref{equ:L_higgs}):
\begin{equation}
 \widetilde{L}^{-} \rightarrow \ell^- Z\;,\; \ell^- H
 \;\;\text{ and }\;\;
 \widetilde{L}^{--} \rightarrow \ell^- W^-\;.
\end{equation}
Note that the decay mode $\widetilde{L}^{-} \rightarrow \nu W^-$ is suppressed. Furthermore, we shall assume that the $\widetilde{Y}$ mediated processes dominate over the electroweak decay
$\widetilde{L}^{--} \rightarrow \widetilde{L}^{-} \pi^-$, which is, strictly speaking, allowed because of the mass splitting of the doublet components due to quantum corrections \cite{Q_split}. As a result, we may treat both $\widetilde{L}^{-}$ and $\widetilde{L}^{--}$ on an equal footing. Consequently, we see that the three relevant states from $pp$ collisions ($\widetilde{L}^{-}  \widetilde{L}^{+}, \widetilde{L}^{--}  \widetilde{L}^{++}$ and $\widetilde{L}^{-}  \widetilde{L}^{++}$) will lead to a generic $\ell^+ \ell^- X X'$ signal, where $X X'$ denotes one of $ZZ, HH, ZH, W^+ W^-, ZW^+$ or $H W^+$.

For instance, the case $pp \rightarrow \gamma, Z \rightarrow \widetilde{L}^{--}  \widetilde{L}^{++}$ will give rise to a $\ell^+W^+\ell^-W^-$ state. Subsequently, the $W$ bosons may decay leptonically or hadronically leading to one of the final states (in order of descending branching fraction): $\ell^+ \ell^-  j$, $\ell^\pm \ell^\pm  \slashed{E}_T \ell^\mp j$ or $\ell^+\ell^-\ell^+\ell^- \slashed{E}_T$, where $\slashed{E}_T$ and $j$ denote the generic missing transverse energy and jets respectively. The relevant SM background processes for this case are
$pp \rightarrow t\overline{t}W^\pm, Z W^+W^-$ (and perhaps $pp\rightarrow t\overline{t}$). Amongst the three final states, $\ell^\pm \ell^\pm \slashed{E}_T\ell^\mp  j$ seems to be the most promising for probing the mass of $\widetilde{M}$ as one may gain information from studying either the two same-sign leptons invariant mass distribution or the invariant mass of the opposite-sign lepton with two jets. 

Another interesting case to consider is the $pp \rightarrow W^+ \rightarrow \widetilde{L}^{-}  \widetilde{L}^{++}$ decay chain. Although there are more possibilities for the intermediate state $\ell^+\ell^- X X'$, if one assumes that any subsequent $W$ boson decays only leptonically, then the final signal one gets is either $\ell^+\ell^+\ell^-\slashed{E}_T$ or $\ell^+\ell^+\ell^-\slashed{E}_T j$. The typical SM background one must confront with here is coming from vector boson decays (e.g. from $pp \rightarrow ZW^+$).
According to a recent analysis in \cite{DelNobile:2009st}, such channels can provide another promising way to search for these exotic particles. Note, however, that  the cleaner state $\ell^+\ell^+\ell^-\slashed{E}_T$ is only possible if $\widetilde{M}$ is low enough. 

\section{Conclusion}

Given that the discovery of nonzero neutrino masses demands an extension to the lepton sector, it is natural to ask what might be the simplest ways that new physics can couple to the SM particles. Concentrating on the lepton sector only, we have followed the approach of \cite{Chua:2010me} and introduced exotic particles into the SM via some ``minimal couplings''.

Since many of the exotic particles defined in this way turn out to be equivalent to those studied in other new physics models, we identify that only a couple of possibilities remain unexplored in the literature. While the work of \cite{Chua:2010me} focused on one of them, in this paper, we have presented the analysis for the last remaining possibility, namely, the exotic doublets $\widetilde{L}$ which couples to the RH charged lepton singlet.

Using a formalism  similar to that presented in \cite{Chua:2010me}, we have defined the key quantity, $\lambda$, which encapsulates the new physics effects caused by the introduction of these exotic $\widetilde{L}$'s. 
In particular, we note that the off-diagonal entries of this $\lambda$ matrix are the origins of any new FCNC phenomenologies. By invoking the limits from low-energy experiments, constraints are then placed on these entries that control the coupling strength to the exotic $\widetilde{L}$'s. Such an investigation can be quite useful given that these minimally coupled particles may give rise to definite collider signatures at the LHC in the future \cite{DelNobile:2009st}.

In this paper, the processes considered include leptonic $Z$ decays, LFV $\ell\rightarrow 3\ell$, $\ell\rightarrow \ell'\gamma$ decays, as well as $\mu$-$e$ conversion in titanium nuclei. The collection of constraints are displayed in Table~\ref{table:all_results}. We have found that  diagonal elements, $|\lambda_{\alpha\alpha}|$, could be as big as \order{10^{-3}} only, while the most stringent bounds for the off-diagonal values are from LFV $\ell\rightarrow 3\ell$ decays except for $|\lambda_{e\mu}|$ which has its strongest bound coming from the $\mu$-$e$ conversion process.
We anticipate that some of these limits will improve significantly when the next generation of experiments have reached their proposed sensitivity.

Finally, the contribution to the lepton anomalous magnetic moment in this model is calculated. The explicit computation of the relevant lowest-order loop graphs shows that any potential contributions is far too small to be detected in experiments at the present time. As a result, introducing this type of exotic doublet of leptons into the SM cannot resolve the muon $g-2$ anomaly.

\section*{Acknowledgements}

The author would like to thank C.~K.~Chua for discussion and suggestion of ideas; C.~H.~Chen, D.~P.~George and R.~R.~Volkas for comments. This work is supported in part by the NSC (grant numbers: NSC-97-2112-M-033-002-MY3, NSC-99-2811-M-033-013 and NSC-100-2811-M-006-019) and in part by the NCTS of Taiwan.






\end{document}